\documentclass[screen,sigplan]{acmart}
\newif\ifpreprint

\preprinttrue



\acmISBN{}
\acmPrice{}
\acmDOI{}
\acmYear{2020}
\copyrightyear{2020}
\acmConference[DLS '20]{Proceedings of the 16th ACM SIGPLAN International Symposium on Dynamic Languages}{November 17, 2020}{Virtual, USA}
\acmBooktitle{Proceedings of the 16th ACM SIGPLAN International Symposium on Dynamic Languages (DLS '20), November 17, 2020, Virtual, USA}

\ifpreprint

\setcopyright{none}

\else

\setcopyright{acmlicensed}
\acmPrice{15.00}
\acmDOI{10.1145/3426422.3426984}
\acmISBN{978-1-4503-8175-8/20/11}
\acmSubmissionID{dls20main-p13-p}

\fi

\graphicspath{{images/}}

\usepackage{booktabs, listings, xspace, xcolor}

\definecolor{Pallete1}{RGB}{27,147,119}
\definecolor{Pallete2}{RGB}{217,95,2}
\definecolor{Pallete3}{RGB}{117,112,179}
\definecolor{Pallete4}{RGB}{27, 158, 119}
\definecolor{Pallete5}{RGB}{217, 95, 2}
\definecolor{Pallete6}{RGB}{117, 112, 179}
\definecolor{Pallete7}{RGB}{102,102,102}

\definecolor{LightGray}{rgb}{.92,.92,.92}
\definecolor{Gray}{rgb}{.3,.3,.3}
\definecolor{DarkGray}{rgb}{.5,.5,.5}

\lstdefinestyle{R}{ %
  language=R,
  morekeywords={assign, delayedAssign, +, in, FALSE, TRUE, warn},
  deletekeywords={count,body, env, equal, runif, trace, readline, args, exp, t, all, <, -, add,
  warnings, max, system, time,list,c,rm,sys,frame,any,is,na, length, double},
  otherkeywords={<-,<<-,!,!=,~,\$,*,\&,+,^,\%,\%/\%,\%*\%,\%\%,>,<,-,/, = },
  breaklines=true
}
\lstdefinestyle{Rin}{ %
  style=R,
  breaklines=false
}
\lstdefinestyle{Cpp}{ %
  language=C++,
}

\newcommand{\rc}[1]{{\lstinline[style=Rin]!#1!}\xspace}

\newcommand{\ie}{\emph{i.e.},\xspace}
\newcommand{\R}{{\textsf{\v{R}}}\xspace}

\newcommand{\aOverhead}{0.6\%\xspace}
\newcommand{\bOverhead}{0.3\%\xspace}
\newcommand{\cOverhead}{2.7\%\xspace}

\newcommand{\aOverThree}{4\xspace}
\newcommand{\bOverThree}{5\xspace}
\newcommand{\cOverThree}{19\xspace}

\newcommand{\aWorstOH}{9.9\%\xspace}
\newcommand{\bWorstOH}{9.1\%\xspace}
\newcommand{\cWorstOH}{12.7\%\xspace}

\begin{document}

\title{Sampling Optimized Code for Type Feedback}
\subtitle{\bigskip}

\author{Olivier Fl{\"{u}}ckiger}
\affiliation{%
  \institution{Norheastern University}
}
\orcid{0000-0003-0512-9607}

\author{Andreas W{\"{a}}lchli}
\affiliation{%
  \institution{University of Bern}
}

\author{Sebasti\'{a}n Krynski}
\affiliation{%
  \institution{Czech Technical University}
}
\affiliation{%
  \institution{National University of Quilmes}
}

\author{Jan Vitek}
\affiliation{%
  \institution{Northeastern University}
}
\affiliation{%
  \institution{Czech Technical University}
}




\begin{abstract}
  To efficiently execute dynamically typed languages, many language implementations
  have adopted a two-tier architecture.
  The first tier aims for low-latency startup times and collects dynamic profiles, such as the
  dynamic types of variables. The second tier
  provides high-throughput using an optimizing compiler that specializes
  code to the recorded type information. If the program
  behavior changes to the point that not previously seen types occur in specialized
  code, that specialized code becomes invalid, it is deoptimized, and control
  is transferred back to the first tier execution engine which will start specializing anew.
  However, if the program behavior becomes more specific, for instance, if a
  polymorphic variable becomes monomorphic, nothing changes. Once the program is running
  optimized code, there are no means to notice that an opportunity for optimization
  has been missed.

  We propose to employ a sampling-based profiler to monitor native code
  without any instrumentation. The absence of instrumentation means that when the profiler
  is not active, no overhead is incurred. We present an implementation
  is in the context of the \R just-in-time, optimizing compiler for the R
  language. Based on the sampled profiles, we are able to detect when the
  native code produced by \R is specialized for stale type
  feedback and recompile it to more type-specific code. We show that sampling
  adds an overhead of less than 3\% in most cases and up to 9\% in few
  cases and that it reliably detects stale type feedback within milliseconds.
\end{abstract}

\maketitle

\ifpreprint
\vfill\eject
\fi

\section{Introduction}\label{sec:intro}

Efficiently executing dynamic languages is a challenging endeavor as
the many articles, conferences and industry resources dedicated to this topic
prove. The hurdles are plentiful, solutions must account for dynamically typed
variables~\citep{gal09}, for values with dynamic object layouts~\citep{cha89},
for late-bindings affecting the call-graph~\citep{pal01},
for introspective and reflective operations~\citep{dub13}, just to name a painful few. Most
successfully adapted to this environment are virtual machines with multi-tier just-in-time compilers.
The idea being to combine different execution engines occupying different points in
the compile-time versus execution-time trade-off space. Early tiers, typically
implemented by interpreters, favor low latency, whereas late tiers favor
throughput and are realized by optimizing native compilers~\citep{fin03a}. To deal with the
aforementioned dynamic nature of the source languages, the virtual machine
monitors program execution and collects profiles to be propagated from one tier
to the next~\citep{hol94a}. This allows the compiler to use information from previous runs to
tailor optimized code of future runs to the observed behavior. Programs are
assumed to behave such that the observed properties stabilize over time and
after a so-called warm up phase the execution reaches a stable peak performance
level. Unfortunately, as noted empirically by \citet{bar17} reality does not necessarily
conform to this model. If the behavior of programs changes over time,
performance can still degrade late in the execution due to a newly emerged
behavior, defying the traditional notion of a clear warmup phase. At the same time,
profiles recorded early in the execution of a program can still affect
performance of a later tier by degrading optimization choices, even if that
profile information is stale. The reason is that the highest tier, reserved for
the hottest functions, does not collect profiling information anymore, to avoid
overheads that would reduce the peak performance. In other words, in this post-warmup
phase, the execution flies blind and the compiler commits to optimizations which,
barring deoptimization, stay fixed for the rest of the execution.
While a VM has guards and deoptimization to detect when types
change, such that past assumptions do not hold anymore, there is typically no
mechanism to detect when new optimization opportunities arise and new
assumptions could be made.

In this article we propose to combine traditional instrumentation based
type feedback, with another well-known technique, namely a
sampling-based profiler. While we use instrumentation to propagate information
from baseline to optimized code, the sampling profiler is used to monitor
values in optimized native code and trigger re-optimization.
With this additional profiler we specifically address the issue
of optimized code being compiled for stale type feedback. The advantage of a
sampling profiler is that it does not rely on instrumentation and only incurs
overhead when it is engaged. In particular, our aim is to detect when the
feedback collected with instrumentation during warmup indicated a more generic
behavior than what is observed after optimizations, in the random samples gathered
later in the execution.
Our implementation currently focuses on sampling the primitive types of boxed
values in optimized native code.
We detect, for instance, variables in optimized code that were
assumed to be more polymorphic than the actual values observed in the samples.
Or, in general, any program location where the type feedback used during
optimization is more generic than the current samples.
We show that for three benchmarks
representing different origins of stale type feedback we identified in \R, automatic 
re-optimization leads to significant peak performance
speedups between 1.1$\times$ and 1.5$\times$. Our evaluation shows the overhead of our
profiler on the \R
benchmark suite to be mostly within 3\% and sometimes up to 9\%, when the profiler is
continuously running. With this overhead, stale type feedback can be detected
within milliseconds. Finally, we discuss
exciting new profiling opportunities by random sampling over instrumentation
that could be explored in future work.

\begin{figure}
  \includegraphics[width=0.9\linewidth]{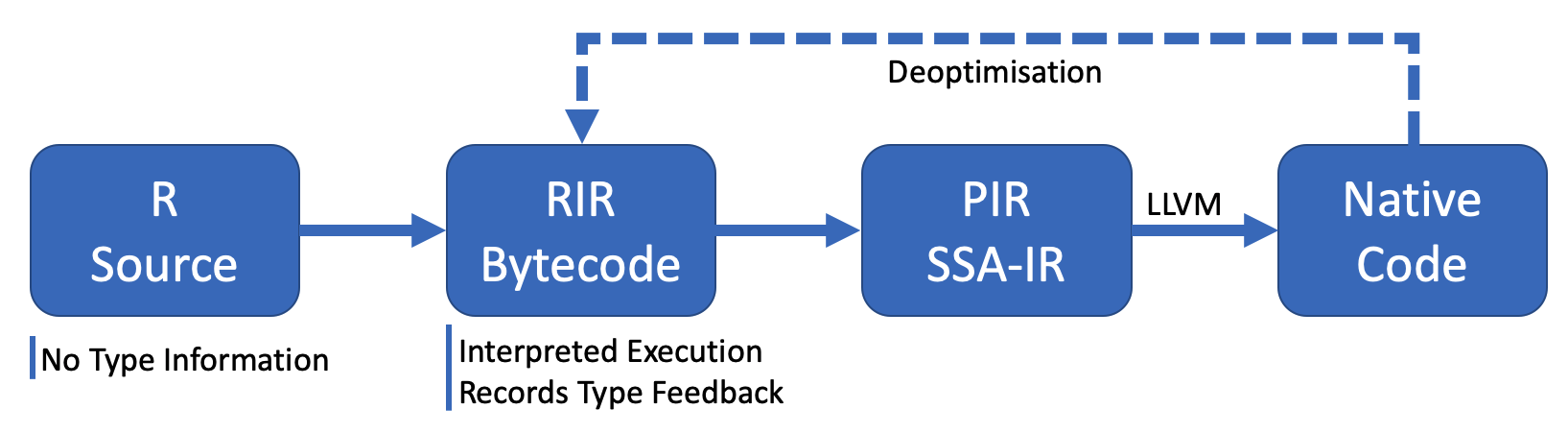}
  \caption{\R compilation pipeline}
  \label{fig:rPipeline}
\end{figure}

We present the following contributions:
\begin{enumerate}
\item A strategy for a posteriori detection of missed speculation opportunities
    using a backup sampling-based profiler monitoring heap values. The profiler
    uses the Performance Monitoring Unit (PMU) as a low-overhead sample trigger
    in an uncooperative environment.
\item An evaluation of this strategy with an implementation on top of the \R
    research virtual machine for R. We report on overhead, stability and
    performance improvements.
\end{enumerate}

The following \autoref{sec:background} provides an introduction to the
difficulties in optimizing R and discusses related work. Then, we present our solution
in \autoref{sec:solution}, an in-depth presentation of the implementation in
\autoref{sec:implementation}, and our evaluation in \autoref{sec:eval}. We discuss
the results and future work in \autoref{sec:discussion} and conclude in
\autoref{sec:conclusion}.

\ifpreprint
\bigskip
\fi

\section{Background}\label{sec:background}

R is a programming language mainly used for statistical and computational data
science applications \citep{rco20}.
The language is notoriously difficult to compile, mainly because of its rich
reflective interface, late binding, lazy evaluation and exposed internals \citep{ecoop12}.
For example, it is possible for a function to modify, add or remove local
variables at runtime.
\R\footnote{https://github.com/reactorlabs/rir} is a relatively new just-in-time compiler
for R which integrates into the reference R implementation \citep{dls19}. As
shown in \autoref{fig:rPipeline}, \R features a
two-tier optimization pipeline with a bytecode called RIR, an optimizing
compiler with a custom intermediate representation called PIR, and an LLVM based
native backend. The RIR interpreter gathers type feedback that is later used for
optimization decisions. To that end there are several recording bytecodes that
can be used to track values, types, call targets and branch targets. This profiling
information is then used to annotate PIR values in the optimizer, and consumed
by speculative optimization passes. The final optimized native functions will
contain type checks to guard for behaviors that fall outside the profiled range.
In case they fail, functions are deoptimized and execution transferred back to
the interpreter.

Since R is dynamically typed, for most values the type of a variable is only known at runtime.
In GNU R values are stored in a structure named \emph{SEXP} for S-expression.
Those contain a header with common information such as the primitive type, the
size, and more attributes that allow, for example, to tag the different object
types of the different object systems in R. All values are vectorized, scalars
are represented by vectors of length one. \R has decided to keep the object
layout of GNU R for compatibility reasons with packages that rely on their
exact layout. This means that all variables are stored boxed on the heap by
default. The IR of the optimizer, PIR, is typed and uses speculative optimizations
to narrow down the static type of values. For instance, numbers which are
speculated (or inferred) to be monomorphic and
scalar will be lowered to unboxed registers in LLVM. For boxed values, \R
maintains a shadow stack of local pointers to make them
reachable by the GNU R garbage collector \citep{hen02}.
As type feedback, \R records the primitive type, whether a value is a
scalar (\ie a vector of length one), eagerness of call-by-need arguments, and
very approximate information about attributes. Type feedback is recorded by a
dedicated RIR instruction which records the type of the value at the top of the
operand stack. The result is an approximation of the union type of all observed
values and can then be extracted to a PIR type used for speculation.

Where does stale type feedback originate from in \R?
One possible origin comes from changing
global state. R is designed to be an interactive environment and there are also
frontends following the notebook paradigm. During such an explorative session, it is common that
some values are interactively changed and a part of the computation executed
again. The traditional idea of collecting profiles during a warmup phase and
then indefinitely committing to a speculation on the future behavior is not well suited 
for live environments.
Another very common problem in R programs are accidentally polymorphic
variables. Consider for instance the following implementation of a counter using
a closure:
\begin{lstlisting}[style=R]
counter <- function() {
  count <- 1L
  function()
    count <<- count + 1
}
\end{lstlisting}
\noindent Here, the variable \rc{count} was intended to be an integer as the
\rc{1L} suffix suggests. However, accidentally on the first increment a floating-point \rc{1}
is added, converting it to double from then on. Therefore, the observed types of
\rc{counter} are integer and double, but dynamically the integer only
occurs up to the first invocation. From then on, the function
might as well be specialized for the double case. 
A third very common case for erroneous feedback is not really a case where the feedback
is stale, but instead a case
where the dynamic analysis is not precise enough and merges unrelated profiles.
A typical situation occurs in polymorphic library code with different callers.
If we consider the following \rc{add} function:
\begin{lstlisting}[style=R]
add <- function(a,b) a+b
add(1,2)
add(c(1,1),c(2,2))
\end{lstlisting}

\noindent As can be seen, this function is called with scalars as well as
vectors. Therefore the profiling information for \rc{add}
records both arguments to be either scalar or vectors.
When this \rc{add} function is inlined, the PIR optimizer imports type feedback
from the inlinee into the caller. Therefore at both call sites \rc{add(1,2)} and
\rc{add(c(1,1),c(2,2))} the merged type feedback of both invocations will be
imported. The problem is that, in general, type feedback lacks dynamic context.
Of course in this simplistic example the actual type of the variables
can be inferred from the static context. But in general this is a problem that can
easily be observed in practice and does have negative effects on performance.
As expressed by \citet{piz18}, it is one of the reasons for the appeal of
multi-tier architectures, e.g. employed in JSCore. The second tier performs inlining, and at the same time
still features instrumentation to
record profiles.
This allows a virtual machine to record context sensitive 
profiles for inlined functions, as for instance \citet{wur12} mention, by resetting profiling after
inlining.

\subsection{Related Work}
Several advances towards using a continuously running profiler with a just-in-time
compiler in what is essentially a feedback-directed optimization loop have been
made:

\citet{wha00} describe a low-overhead sampling profiler for Jalape\~{n}o, a Java virtual
machine. Their profiler is implemented with a busy loop in a dedicated thread and is able to
capture profiles concurrently to the user program threads execution.
They also present an efficient data structure to record context-sensitive profiling
information. As described by \citet{bur99}, in Jalape\~{n}o optimization
decisions are primarily directed by the sampling-based profiling.
The approach is further refined by \citet{sug11} by enriching the profiling with
instrumentation that is dynamically added and removed through recompilation.
In that sense our approach has the inverse priorities. We rely mainly on
traditional instrumentation already
present in \R, and only engage the sampling approach to detect changing behavior
over time that would not be visible otherwise.

\citet{arn12} present a profiler for the Jikes RVM that performs probabilistic
sampling using
instrumentation. The advantage being that the approach is portable and does
not rely on operating system or hardware support. The profile information
focuses on low-level control flow, such as edge counters, and is
used for optimizations such as splitting, reordering, or unrolling.

Another approach for gathering probabilistic samples without instrumentation is
to use hardware performance monitors by the CPU. \citet{sch07} query the
performance counters on cache misses and use that information to improve
allocation co-locality in a virtual machine for Java.
\citet{kis03} propose techniques for continuous program optimization at the
native code level. Their system instruments compiled binaries at load time. They
are able to improve cache usage and instruction-level parallelism.

\section{Detecting Stale Type Feedback}\label{sec:solution}

\begin{figure}
\begin{lstlisting}[%
  caption={Example of Feedback Pollution},
  label={lst:pollution},
  style=R]
f <- function() x+x+x+x+1L
if (POLLUTE) {
  x <- 1
  f(); f()
}
x <- 1L
for (i in 1:1000000) x <- x+f()
\end{lstlisting}
\bigskip
\begin{lstlisting}[
  caption={Byte code of function f},
  label={lst:pollution-rir}]
 0   ldvar  x
 9   [double(s), integer(s)]
14   ldvar  x
23   add
     ...
\end{lstlisting}
\end{figure}

To understand how our sampling profiler is integrated into \R, let us introduce a
concrete benchmark for measuring the impact of stale type feedback. In the
following \autoref{lst:pollution}, the flag \rc{POLLUTE} causes the function \rc{f} to be invoked
twice with the global variable \rc{x} being a double. All subsequent invocations
are on integers.
When the \rc{POLLUTE} flag is unset, then this snippet executes $25\%$
faster in \R, due to the variable \rc{x} being fully unboxed in \rc{f}. If on
the other hand, the \rc{POLLUTE} flag is set, the function \rc{f} compiles to 
the RIR code in \autoref{lst:pollution-rir}.
At offset 9 there is a type recording instruction and it has recorded
the top of the operand stack to be either a double scalar or an integer scalar.
When this function is optimized in PIR, the variable \rc{x} loaded from the
global environment is speculated to be an integer or a double. The optimized
function is also annotated with a map as metadata, associating slots in the shadow
stack (which are used by native code for storing boxed values), with original type feedback
locations. In this case after a number of iterations the map contains the
following entries:

\medskip
\begin{lstlisting}[]
- #2->9 : [int(s)] (4), [dbl(s), int(s)]
- #3->9 : [int(s)] (4), [dbl(s), int(s)]
- #4->29: [int(s)] (2), [dbl(s), int(s)]
- #7->81: [<?>]    (0), [dbl(s), int(s)]
\end{lstlisting}
\medskip

\noindent For instance, the first row reads as follows: In the second slot of the
shadow stack, we find a value that corresponds to a type feedback that was
recorded at bytecode offset 9 (see previous listing). The profile of this value
indicates that it is a
scalar integer, we have collected 4 samples and the previous feedback was double
or integer. As can be seen in this example, during optimizations it is possible for a value
to appear multiple times (in which case said values will be aggregated when taking decisions),
or to be absent from the table if it cannot be
traced back to the original type feedback. For some slots we will be able to collect
many samples; for others, the samples will be sparse or even completely missing. 
The goal of the profiler is to detect when a function has many slots with
several samples where the profiled types differ substantially from the previous one.
In this case, the function is marked for recompilation. Recompilation always
starts from the original source function, however, any sampled profile entry with
enough samples overrides the original type feedback.
This enables the optimizer to then ignore the stale, or wrong type feedback that
we collected in the initial run and produce a new function with better peak
performance.

As a concrete example, consider an R user trying to compute values produced by a linear
congruential generator, implemented as follows:

\medskip
\begin{lstlisting}[style=R,xleftmargin=0.4\parindent]
lcg <- function(n)
  for (i in 1:n)
    state <<- (state * 48271) %% 0x7fffffff
\end{lstlisting}
\medskip

\noindent Say the user is interested in the billionth value after an
initial seed of \rc{1}, \rc{2} and \rc{3}. So they will run the following queries
in the REPL:

\begin{lstlisting}[style=R]
> state <- c(1,2,3)
> lcg(1e7)
> state
[1] 1901417813 1655351979 1409286145
\end{lstlisting}
\medskip

\noindent As can be seen here, R is a vectorized language. Therefore, multiple
instances of this pseudo random generator can be run in parallel. As the language
also encourages an interactive exploration style, the user might continue to
query different values for \rc{lcg}, using differently sized vectors for
\rc{state}. Assume we record the run times in seconds of each invocation of
\rc{lcg} during an interactive session in \R: 

\bigskip
\includegraphics[width=0.9\linewidth]{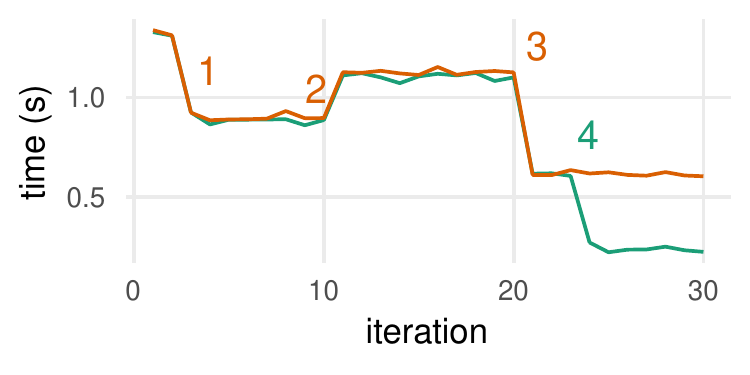}
\medskip

\noindent Let us focus on a number of interesting change points in that graph. First at
(1) the \rc{lcg} function is compiled and execution speeds up by a factor of
1.3. Then at (2) the user switches from triples to vectors with 6 entries, the
execution takes longer again, but not twice as much, thanks to the built in
vectorization. At point (3) the user starts querying \rc{lcg} with a scalar
\rc{state} and performance improves again. Since scalar vectors are so
common in R, the \R compiler has special support for them and can treat them as
machine floating-point numbers. However the stale type feedback from earlier
runs has caused the \rc{lcg} to be optimized to the least upper bound of all
observed types so far, which is a vector of numbers. At (4) we see the two
lines to start differing. The one staying flat is our baseline, the red line
dropping below includes our sampling profiler. The profiler has detected that
\rc{state} is now scalar and that \rc{lcg} should be reoptimized with that
information. The subsequent specialization leads to an over 2
times faster execution.

\section{Implementation}\label{sec:implementation}

This section is dedicated to implementation considerations. Some of the problems
encountered are specific to having to integrate the profiler with an existing
system, others are more generic.
We will start with
the way we trigger the profiler to records samples. Next, we will cover how these
samples are recorded and
what they contain. Finally, we will explain how we decide when to recompile a function and how
the recorded data is used by the compiler in order to improve the compiler's output.

\subsection{Sample Triggering}
The main goal of the sample triggering mechanism is to be performance
neutral when disabled, to record uniformly distributed samples, and to be compatible
with the existing infrastructure. Since GNU R and hence also \R are
single-threaded, it would be difficult to asynchronously collect the
results. Therefore, our implementation relies on interrupting the program at
regular intervals. An obvious approach consists of a separate thread or a POSIX timer interrupting
the program using signals. The issue with said strategy is that interruption can
occur during the execution of a syscall (\ie when the process is executing in
kernel mode) and, in particular, blocking syscalls are aborted or restarted by signals.
Since R packages can contain arbitrary C code, we have no
control over when and how syscalls are executed. We must therefore avoid triggering signals
during syscalls, which is not possible with a timer based approach.

An alternative mechanism for triggering repeating signals is
to use the Performance Monitoring Unit (PMU) present on
modern Intel processors. The PMU allows monitoring of a multitude of information on the performance
of a processor, from counting the number of retired instructions, up to recording
a dynamic call graph. The PMU was first introduced to Intel
Pentium processors in 1993 and expanded in capabilities over time. All Intel Core and Intel Xeon since
the Nehalem architecture (released in 2008), and all Intel Atom processors provide a very capable PMU
with most modern-day features supported \citep{int19}. Other processor manufacturers offer
similar capabilities. For example, AMD Zen processors provide Performance Monitor
Counters (PMC) \citep{amd17}
and some ARM processors also provide similar capabilities \citep{arm14}. In the following sections we
will simply refer to this hardware as the PMU as the work presented in this
paper focuses on Intel processors running Linux. The ideas should however be transferable
to other CPU architectures.

In general, PMU configuration options in Linux can be categorized into either \emph{counting} or
\emph{sampling} mode. In counting mode, the PMU counts a configurable number of
simple events occurring in the CPU, such as retired instructions or cache misses
\citep{moo02,amm97}. The data is kept by counters in the CPU and retrieved by
polling. On the other hand, sampling mode provides access to more complex metrics. The main difference
is that the recorded data must not be polled; instead, the data is written
into a buffer in memory. There is a configurable \emph{sampling period} specifying
a threshold for a particular counter after which a sample is taken and written
to said buffer. Additionally, this configuration mode provides the option to
trigger a signal whenever new data is available.

To use the PMU for sample triggering we configure it in sampling mode and set a
sampling period for the retired instructions. While this number is
irrelevant to us, we use it to
force a change in the sample buffer which in turn triggers a signal to
interrupt the running program. The key feature we need for our profiler is that
it allows us to configure filters such that events in kernel or hypervisor
mode are ignored. This in turn prevents the counter of retired instructions to
be incremented during syscalls, which also prevents the program from being interrupted
at that point \citep{zap09}. The actual sampling is not done by the PMU, but
implemented in a dedicated value profiler. The main ingredients in our PMU
configuration are a PMU configuration to trigger a \emph{perf} event at regular
intervals (but not in kernel mode), an event handler on the perf event 
to raise a signal, and a signal handler
that starts the actual profiling. The perf event is created and linked with a
signal by:
\medskip
\begin{lstlisting}[language=C,
  basicstyle=\small\ttfamily]
// PMU configuration
struct perf_event_attr pe;
memset(&pe, 0, sizeof(pe));
pe.type = PERF_TYPE_HARDWARE;
pe.config = PERF_COUNT_HW_INSTRUCTIONS;
pe.sample_type = PERF_SAMPLE_IP;
pe.sample_period = 1000000;
pe.exclude_kernel = 1;
pe.precise_ip = 3;
// Setup PMU
int fd = perf_event_open(&pe,0,-1,-1,0);
// Connect perf event fd to signal
fcntl(fd, F_SETFL, O_NONBLOCK | FASYNC);
fcntl(fd, F_SETSIG, signal);
fcntl(fd, F_SETOWN, getpid());
// Setup signal handler
...
// Reset events and enable signaling
ioctl(fd, PERF_EVENT_IOC_RESET, 0);
ioctl(fd, PERF_EVENT_IOC_REFRESH, -1);
\end{lstlisting}
\medskip
The full listing for the PMU configuration is included in \autoref{sec:appendix}.

There are some issues with this approach. Depending on the kernel
settings, PMU access is disabled by default for security reasons to prevent
unprivileged users from attaching the PMU to other processes. Moreover, interrupting
execution using signals is not appropriate in all contexts. For instance, it is
not applicable for VMs which allow being embedded into other processes.
Therefore, this part of our approach is \R specific and for other systems it might
be possible to choose a simpler strategy.

\subsection{Sample Collection}

To keep a low overhead and since we are mainly interested in profiling hot
functions, the profiler only inspects the top-most frame. Since our triggering
mechanism invokes it at arbitrary points in the execution, it is not guaranteed
that the function currently being executed is indeed a function produced by the
native backend; it could also be interpreted bytecodes, or some other library, or user
code. In order to detect native code, a reference to the closure object of the
native function is pushed as a marker value at the top of the current frame in
the shadow stack upon entry. This solves two issues at the same time. First, the profiler
is able to detect when a
native function is currently executing. Second, it gains access to the
aforementioned metadata
map and to the backing store for the sampled type feedback (both stored in the
closure object).

The native backend in \R performs unboxing of monomorphic scalar integers or
floating-point numbers. All other values are stored boxed on the heap.
Boxed values need to be reachable by the garbage collector to accurately track
their liveness, which is implemented by keeping them on a shadow stack. To be
able to inspect values in native code, it is therefore sufficient to produce an
accurate mapping of stack slots back to their source value in the backend.
There is a certain disconnect between the original RIR code and the resulting
native code. RIR contains many instrumented locations where values are
sampled for their type. In the optimizer only part of this information is used;
values might be eliminated or duplicated during optimizations. To address these
issues, any type feedback annotation in PIR carries metadata about the
originating location of the feedback. This enables the backend to reconstruct a
mapping from shadow stack slot to the originating position in the RIR bytecode
stream. It is possible that multiple locations map to the same original
position, and most likely many type feedback origins are not present at all
after optimizations. The mapping must account for different stack contents at
different program counter positions. At the time of writing we avoid this
complication by preventing the backend from sharing slots we wish to sample,
thus keeping the mapping static for the whole function.

Once sufficient data has been recorded we might want to trigger the recompilation of
a function. For this decision we take into consideration how many samples we
collected, how many slots we have samples for and, most importantly, how different
those samples are with regards to the original type feedback that was used for
optimization. The threshold we use for our experiments is to trigger
recompilation if more than half of the slots are not equal to the previously
used feedback. To ensure the samples themselves are not stale, the recorded
results are cleared periodically. Recompilation happens before the next
invocation, since \R does not feature a concurrent compiler. On recompilation,
the optimized function is discarded and optimizations start anew using
the source version. However, type feedback with a high
sample count (\ie high confidence on their accuracy) overrides any type feedback
from the original source version.

There are several issues with this simplistic approach. For instance, a program
featuring phase changes which are slightly above the recompilation threshold
will trigger recompilations, which will then be swiftly invalidated.
To use the sampling profiler in practice it will be necessary to fine tune the
heuristics. For example, it should take into account past deoptimizations indicating
incomplete samples by the profiler, or adaptive sampling rates to cater to
different kinds of functions.

\section{Evaluation}\label{sec:eval}

To evaluate our solution we investigate four facets of our implementation,
each of these targets one aspect of the profiler. The first evaluation measures
the overhead introduced by merely running the profiler. The second, attempts to find a
sensible re-optimization threshold. The third evaluation highlights the
examples where our profiles produce useful performance improvements. Finally, we estimate
how common it is for \R optimized code to be specialized for less generic types than what is actually
observed at runtime.

The \R{} benchmark suite used in this paper consists of 46 programs that range from
solutions to small algorithmic problems, and real-world code.
Some programs are variants; they use 
different implementations to solve the same problem.
The suite and \R baseline performance is discussed in detail by
\citet{contextDispatch}.
This pre-existing suite was also expanded and three benchmarks were added to show
potential performance improvements to be gained through use of the profiler.

To deal with warmup phases of the virtual machine 
(\ie iterations of a benchmark during which compilation events dominate
performance), we run each benchmark fifteen times in the same process and
discard the first five iterations. Remaining data is aggregated using mean. We ran experiments on a dedicated
i7-6700K CPU, clocked at 4 GHz, stepping 3, microcode version
0xd6, with 32 GB of RAM and Ubuntu Bionic on Linux kernel version 4.15.0-88.

Our work is an extension to an
open-source virtual machine, available at
{\href{https://r-vm.net}{\v{r}-vm.net}}.  We publish an
executable artifact to reproduce the results~\citep{flu20-artifact-dls}.

\subsection{Overhead Evaluation}\label{sec:eval-overhead}

\begin{figure*}
  \includegraphics[width=0.85\linewidth]{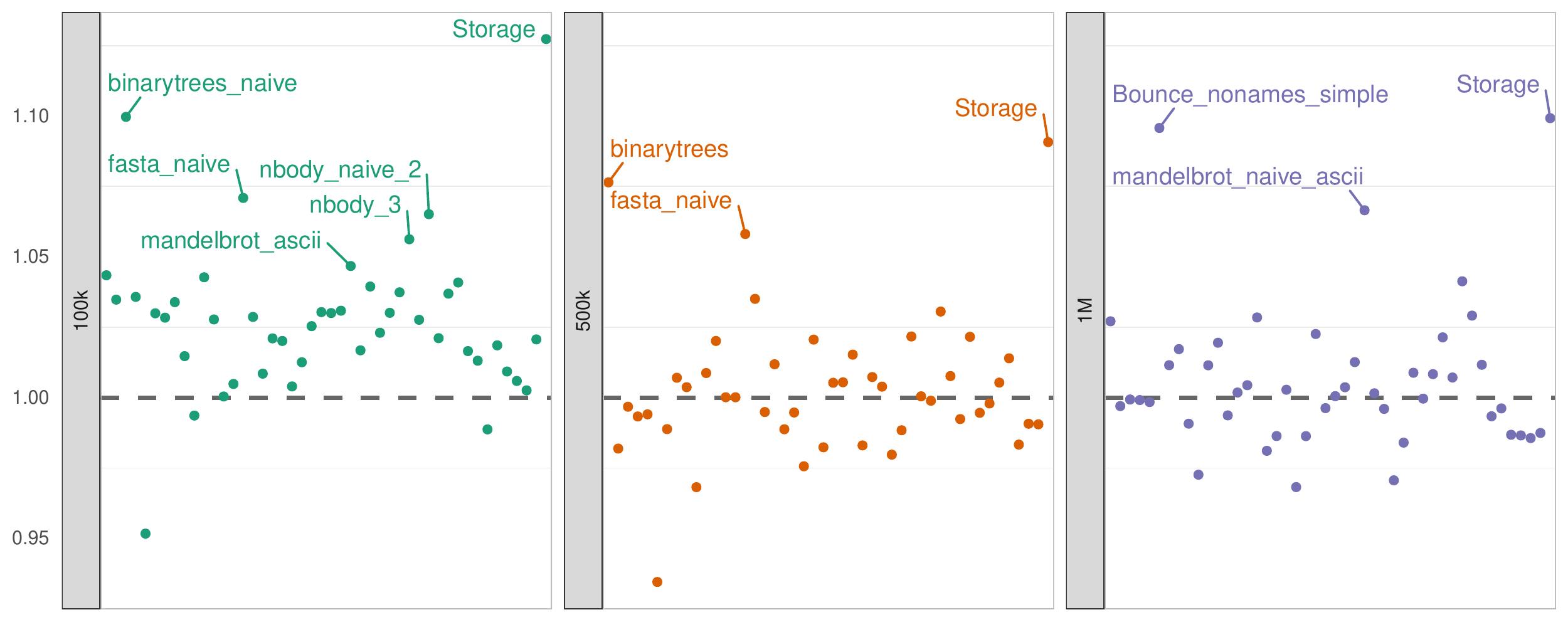}
  \caption{Profiling Overhead per Benchmark}
  \label{fig:eval_overhead}
  \medskip
\end{figure*}

This first evaluation was used to measure the pure performance overhead introduced
by running the profiler. To be able to perform these measurements we modified
the profiler such that while it continues to record type information, that information
is never used in the compiler and a recompilation is never triggered. Doing so, we
avoid measuring potential compiler overhead caused by the profiler triggering a recompilation.
The memory requirements are negligible, as the few recorded slots each only
require 64 bits of additional storage for the metadata, which is very small compared
to the size of the native code itself.

The measurements were performed with four different configurations: The first was a baseline run
with the profiler completely disabled. This was followed by three runs with the profiler
enabled with different sampling periods: One triggered a sample every 100,000 instructions,
one with a sample every 500,000 instructions and the last one with a sample every
1 million instructions. For each benchmark in these runs the median run times
were normalized against those of the baseline run.
While the data is somewhat noisy, we can still gather important information. In some configurations we see
apparent improvements compared to the baseline values. These can be attributed to the inherent noisiness
of the benchmark environment as well as some additional noise introduced through the randomized sampling
locations hit by the profiler and do not indicate actual reproducible performance improvements.

As is clearly visible in \autoref{fig:eval_overhead}, with a sampling period of 1 million instructions
the overhead is minimal. Most benchmarks had a slowdown
of less than 3\% compared to the baseline. Only \aOverThree benchmarks had a slowdown above 3\%. The worst-performing benchmark was \emph{Storage} with a slowdown of \aWorstOH. The mean slowdown over all benchmarks was just \aOverhead.
With a sampling period of 500,000 instructions we observed very similar behavior as before, with only \bOverThree benchmarks showing a slowdown above 3\%. The worst-performing 
benchmark was, once again, \emph{Storage} with a slowdown of \bWorstOH. The mean slowdown was \bOverhead.
Using a sampling period of 100,000 instructions we see a significant slowdown compared
to the previous configurations. Now, \cOverThree benchmarks showed a slowdown of more than 3\% and \emph{Storage} even exceeded 10\% slowdown at \cWorstOH. The mean slowdown was \cOverhead.

\subsection{Threshold Configuration}\label{sec:thresh}
\begin{figure}
  \includegraphics[width=0.95\linewidth]{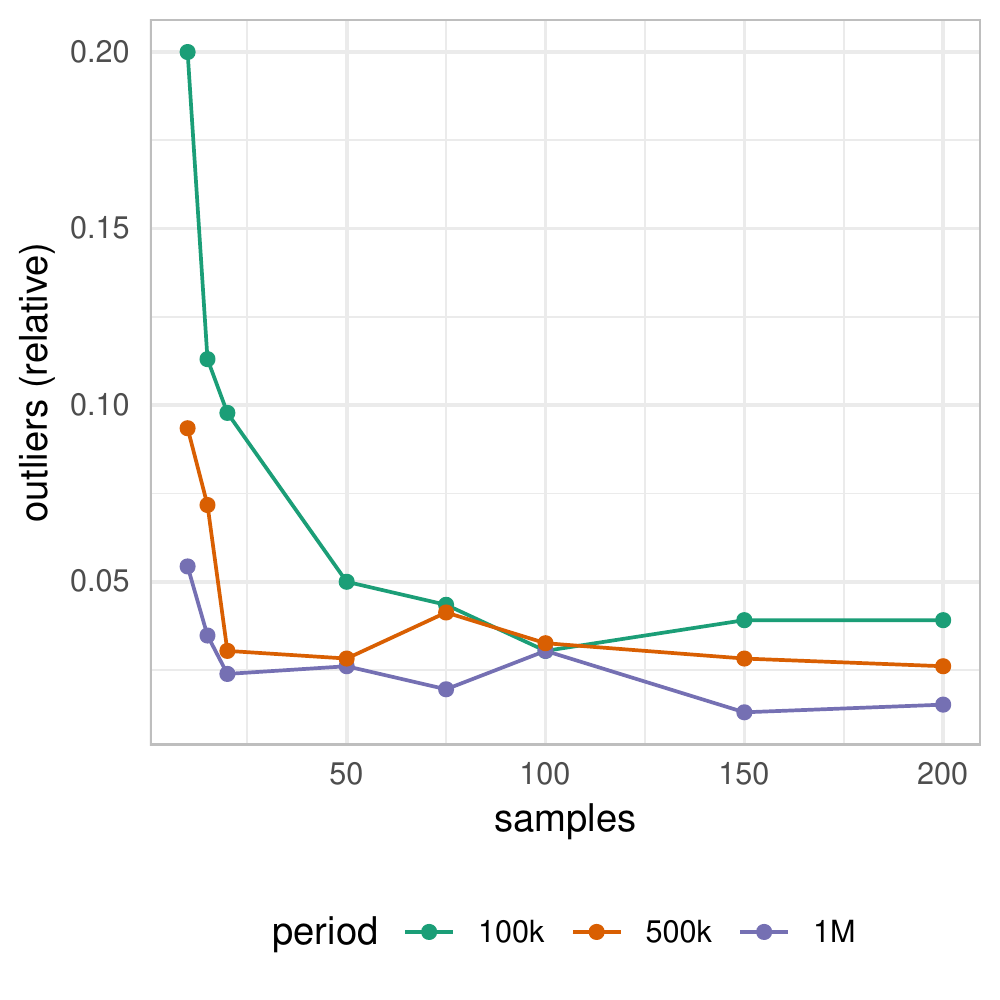}
  \caption{Outlier Frequencies}
  \label{fig:eval_miss}
\end{figure}

This second part of the evaluation was designed to find sensible recompilation thresholds for
the different sample period configurations: 
Low recompilation thresholds allow for quick action by the profiler. This can improve performance in situations
where the resulting compilation produces a good result. But with too low a threshold we introduce a higher
risk of having incomplete data at our disposal when the profiler decides to trigger a recompilation. This can
lead to the newly compiled version being incompatible with what actually is required at that time. This will
invariably lead to subsequent deoptimization events. At too low thresholds this behavior can even become cyclic
where the profiler repeatedly tries to step in and trigger recompilations but never manages to gather a complete
picture before this happens. In such a situation we could observe extreme performance impacts.
An obvious approach to preventing this situation is to use high thresholds. But while this helps
to reduce detrimental recompilations it also delays useful ones. This reduces the performance gains
we can expect in cases where the profiler can theoretically improve performance.
We need to find thresholds that are large enough to avoid having too many detrimental
recompilations while still keeping it as small as possible in order to maximize potential benefits.

We started with a very low threshold of 10 samples and raised
it until significant performance impacts were no longer observed. This was done for the three sampling periods
of 1 million, 500,000 and 100,000 instructions individually. The main difference in profiler configuration compared to
the first part of the evaluation is that now the profiler was allowed to trigger recompilation and the compiler
was allowed to use data collected by the profiler during compilation.

The collected data was then compared to the overhead measurements gathered in \autoref{sec:eval-overhead}.
Each run that had a slow-down of more than 10\% compared to the median run time of the overhead
measurements was flagged as an outlier. The total number of these outliers was counted and reported as
a fractional value relative to the total number of runs recorded. \autoref{fig:eval_miss} shows
these outlier frequencies for different thresholds. In general for the same threshold the higher sampling rate
produced more outliers. We can likely attribute that to the fact that with higher sampling rates we effectively
shorten the time required to reach the recompilation threshold. This shortened sampling time increases the risk
of premature recompilation. Such premature recompilations lead to deoptimization events soon after. When this
happens often enough it is even possible to have a recompilation and a deoptimization event in each run. This
significantly impacts performance.

With a sampling period of 1 million instructions we observed 25 (5.4\%) outliers with a threshold of 10 samples.
This dropped to 12 (2.6\%) outliers for a threshold of 50 samples. Increasing the threshold further does not
seem to improve things: At 100 samples we recorded 14 (3\%) outliers. In fact, starting at a threshold of 20 no significant changes can be observed anymore. 

In addition to the overall observations we can look at specific benchmarks for more information:
The \emph{nbody\_naive\_2} benchmark, for example, had a mean run time of 290
seconds (due to a deoptimization, re-optimization loop) and a median run time of 5 seconds with a
threshold of 10 samples. These values dropped to 1 second each when using a
threshold of 20 samples. This
shows both a significant reduction in run time and increased consistency: With a threshold of 20 the closely
matched mean and median indicate the absence of significant spikes. Of course, the 
extreme outliers in the lower threshold must be mitigated by back-off strategies in practice.

There are, however, a few benchmarks where the increased threshold actually reduced the measured performance:
For example, the \emph{reversecomplement\_naive} benchmark had a mean runtime of 323ms at a 
threshold of 10. It increased to 393ms at a threshold of 20. This is attributable to the fact that,
at larger thresholds, a recompilation event
is measured in the ten benchmark runs that were hidden inside the warmup phase for smaller thresholds.

Overall, for a sampling period of 1 million instructions and thresholds starting at 20, we observed exactly one slow run that significantly increased mean run times.
The median run times however remained essentially unchanged.

When considering a sampling period of 500,000 instructions, the behavior is very similar as before: While at a threshold of 10 we 
observed 43 (9.3\%) outliers, this dropped to just 14 (3\%) at a threshold of 20. And above 20 samples we observed basically identical outlier counts as with a sampling period of 1 million instructions.

With a sampling period of 100,000 instructions we see a high outlier count
of 92 (20\%) for a threshold of 10.
This drops down to 20 (4.3\%) at a threshold of 75. At a threshold of 10 it behaves clearly worse than with a sampling period of 1 million
 instructions. Due to the significantly lower sampling period we will collect samples quicker.
 When using the same threshold we effectively make the profiler more eager
 in its optimizations since it only needs to measure over a shorter time frame.

\subsection{Performance Improvements}\label{sec:eval-improve}

\begin{figure*}
  \includegraphics[width=0.75\linewidth]{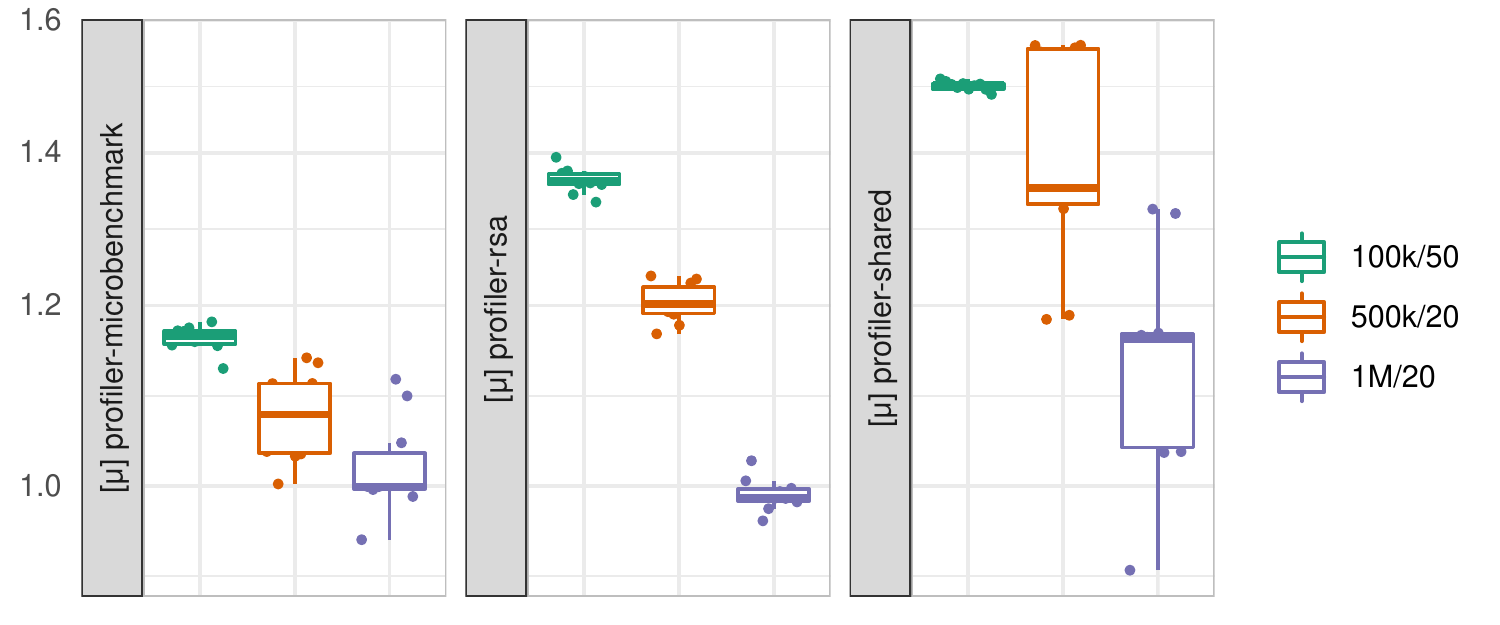}
  \caption{Performance Improvements}
  \label{fig:eval_improve}
  \medskip
\end{figure*}

The third part of the evaluation is designed to show potential performance improvements gained through
use of the profiler. To that end we added three additional benchmarks to the \R{} benchmark suite:
\emph{profiler\_microbenchmark}, \emph{profiler\_rsa} and \emph{profiler\_shared}.

The \emph{profiler\_microbenchmark} is essentially the same as the example function \rc{f} in \autoref{sec:solution}.
It is designed to test a situation where a single stale type was introduced before compiling the function.
Here we start with a floating-point number. But after two calls we change to an integer. The function
will initially be specialized to both integer and floating-point numbers. The profiler should detect
that only integers are present from now on and should at some point trigger a
re-optimization. If the pollution is statically removed, then \R executes the
function about twice as fast.

\begin{figure}
\begin{lstlisting}[
  style=R, caption={Naive RSA implementation}, label={lst:rsa}
]
p1 <- 971
p2 <- 383
n1 <- p1 * p2       # float
e <- 17
encrypt <- function(msg) {
  p <- 1
  a1 <- msg
  for(i in 1:e) {
    p <- p*a1
    p <- p%%n1
  }
  p
}
for(i in 1:n) encrypt(i)
n1 <- 371893L       # integer
for(i in 1:n) encrypt(i)
\end{lstlisting}
\end{figure}

The \emph{profiler\_rsa} benchmark shown in \autoref{lst:rsa} contains a simple RSA encryption
implementation with small key numbers.
In this function we start with the key element \rc{n1} as a floating-point number
(calculated from \rc{p1} and \rc{p2}).
The function is repeatedly called and is therefore compiled expecting floating-point numbers in \rc{n1}.
At some point, however, \rc{n1} changes and is set to an integer. After that, the function is again called
several times. This will lead to a deoptimization followed by compilation for
floating-point \emph{and} integers.
At this point the profiler is supposed to step in and detect the stale feedback
for floating-point and cause function \rc{encrypt} to be recompiled
to only support integers in \rc{n1}. This benchmark shall serve as an example for a program with
a phase change: After a stable behavior in a first phase, some datatypes change as the program transitions
into its second phase. For a monomorphic \rc{n1} the function executes about twice as fast. This
is due to \R not yet supporting unboxed values that can be either integers or
floating-point values, since the resulting dynamic type conversions would be
difficult to track.

\begin{figure}
\begin{lstlisting}[style=R, caption={Trivial functions polluted by multiple callers}, label={lst:shared}]
id <- function(a) {
  # prevent inlining
  while (F) a;
  while (F) a;
  a
}
add <- function(a, b) id(a)+id(b)

poison=structure(1, class="foo")
add(poison, poison)

test <- function() {
  s = 0
  for (num in 1:500000) {
    s = add(s, num)
  }
  s
}
for (i in 1:n)  test()
\end{lstlisting}
\end{figure}

The third benchmark \emph{profiler\_shared} shown in \autoref{lst:shared}
features trivial \rc{id} and \rc{add} functions with two different callers.
Initially the \rc{add} function is called with a number that has a class
attribute attached to it, namely with the argument \rc{poison=structure(1, class="foo")}.
Because \rc{poison} is an object and \rc{add} does not use it directly, it is passed as a promise
to \rc{id} where it is in fact evaluated. The \rc{id} function is also written in such a way that it will not be inlined.
This prevents \rc{add} from observing that \rc{poison} is in fact a number. In the following part
of the benchmark,
the \rc{test} function is called repeatedly which, in turn, repeatedly calls the \rc{add} function with simple numbers.
The \rc{add} function is inlined into \rc{test} importing with it the unrelated
type feedback of the earlier calls with \rc{poison} as arguments.
We lose performance improvements we could have had if \rc{add}
had the information
that it was dealing with normal numbers. This is, however, a fact the profiler should
be able to detect and trigger reoptimization for. This is
a second example of a scenario where previously relevant type information
is no longer needed. But other than
in the \emph{profiler\_microbenchmark} this time it is the combination of an
object and a number type instead
of two number types. And instead of a global variable that changes,
it is the method's parameter in a situation where the function would
normally not be specialized. By statically removing the \rc{add(poison, poison)}
calls, \R is able to execute this function six times faster.

\paragraph{Results}

For the evaluation we used the threshold values determined in \autoref{sec:thresh}.
For each sampling period configuration, we used the smallest measured threshold
above which no significant improvements in the outlier count could be observed. This is
20 for periods of 1 million and 500,000 instructions and 75 for 100,000 instructions.
Tuning the compilation threshold is not easy and there is a clear
trade-off: These benchmarks take about 1 to 2 seconds to complete.
With a lower threshold, optimizations can be applied much quicker
but we risk overly eager optimization. With a higher threshold, we reduce the risk
of overly eager optimization but delay optimization where they would make sense.
The \rc{microbenchmark} and the \rc{rsa} benchmark could be improved by at most
2$\times$, the \rc{shared} benchmark by 6$\times$. The actual improvements
are more modest for three reasons: First, it takes the profiler some time to
detect the phase change. Second, the benchmark includes re-compilation
time. Lastly, the theoretical limit was determined by statically changing the
benchmarks to exhibit only the stable behavior, whereas the profiler must use
speculative optimizations with run-time guards.

\autoref{fig:eval_improve} shows that for all three benchmarks, the 1 million instruction
sampling period performed the worst of all three. In fact, for each benchmark, the smaller
sampling periods are always performing better. This makes sense, since the difference in
thresholds is smaller than the difference in the sampling period.
This leads to shorter sampling times and quicker reoptimization when using smaller sampling periods.

We also observe a sampling period of 1 million instructions and a threshold of
20 to not yield any substantial performance improvements. In all three benchmarks
there are runs that show a reduction in performance. However, with a sampling period of 500,000
instructions at the same threshold, we are capable of reoptimization soon enough that
we see a significant increase in performance in all three benchmarks.

While using a sampling period of 100,000 instructions with a threshold of 50 yields
better performance, it is also the configuration that produces the most outliers
in \autoref{sec:thresh} of the configurations presented here.
This once more highlights the trade-off present in tuning the profiler for
a specific application. To conclude, taking into account both the outliers produced
in \autoref{sec:thresh} and the benefit gained in these three examples, we
consider a sampling period of 500,000 instructions and a threshold of 20
to be the best combination.


\subsection{Estimation of Impact}

To estimate how common it is for R code optimized by \R to feature types which
are not as narrow as actually observed at run time, we performed an additional
experiment on the benchmark suite using a modified version of \R. The
backend was changed to instrument the native code such that every boxed value
produced is recorded according to the same approach as the sampling profiler
would. This allows us to simulate a maximally accurate profiling run. During
the execution of the benchmark suite we observed that 36 out of 46 benchmark
programs contained at least one instance of optimized code encountering less
generic value types, mostly during warmup (i.e. before inlining).

\begin{figure}
\begin{center}
\small
\begin{tabular}{ l l l l }
           & narrower & changed & optimizable \\
  warmup  (mean)    & 4   & 21 & 13   \\
  warmup  (median)  & 1   & 1  &  0   \\
  stable  (mean)    & 0.6 & 6  & 3.5  \\
  stable  (median)  & 0   & 0  & 0    \\
\end{tabular}
\end{center}
  \caption{Optimization Opportunities}
  \label{fig:opportunities}
\end{figure}

However, as can be seen in \autoref{fig:opportunities}, after the
warmup phase
\R does not under-speculate much,
as expected since it is optimized for this suite.
If it does, the missing speculation would not always lead to improvements, since
there might not be a target
optimization. The row \emph{narrower} indicates that the profiled type is a
subtype of the original type feedback, and \emph{changed} indicates that it is a
different, unrelated type. We conservatively define narrower or changed types to be an
optimization opportunity (labeled \emph{optimizable}),
if they would lead to either unboxing, or optimizations through ruling out dynamic
dispatch. Some benchmarks contain a large number of recorded optimization
opportunities, since one missed opportunity likely occurs multiple times if it
concerns the same variable.

\section{Discussion and Future Work}\label{sec:discussion}

There are several open questions which are not answered yet in this work. The
main question of course being, whether the technique is successful at improving the
performance of real world R programs. We believe to answer this question
our implementation has to be extended to include more properties and provide a
slightly more holistic notion of \emph{missed optimization opportunities}.
Moreover, the evaluation we performed of the current implementation
led us to the realization that the traditional benchmarking methodology has
limitations that prevent us from measuring any performance gains of such a system.
The existing \R benchmark suite is designed to measure peak performance. The approach
involves repeatedly invoking the same function and only measure
the run time once it stabilized. Each in-process iteration of one benchmark
performs the same sequence of operations from start to finish. Contrasting this
with the goal of profile guided re-optimization, we notice that any profile
information that appears stale within one iteration of the benchmark, is
guaranteed to repeat again when the next iteration is invoked.
Not surprising we do not measure any performance improvement, since
every iteration does exactly the same thing; if the profiler triggers some
re-optimization, we are guaranteed to cause a deoptimization event on the next
in-process iteration, since we always start again from the same state. To be
able to measure a positive
effect within this framework we are therefore presented with two equally
unattractive choices: Either we keep some global state across iterations and
cause the first iteration to be different from all subsequent ones, or we
reset all compiled code at the beginning of each iteration (or similarly, measure
whole-process iterations instead of in-process iterations). The former approach
does not allow us to measure how fast the profiler is at engaging the compiler
and causing the performance to improve, since all of the re-optimization happens
in the first
few iterations and which we will discard as warmup runs. With the latter approach,
taken in \autoref{sec:eval-improve},
we simulate the whole process from warm up, then optimization with stale
feedback, then re-optimization, for every measured iteration. This means our
reported numbers include compilation time, profiling time, time spent in the
interpreter, and so on. Such a measurement includes much more noise from
unrelated causes and therefore we need to take adequate precautions to correctly interpret
the numbers. For instance, the benchmark needs to be such that the majority of
the time is spent in the final, stable, re-optimized native code. If not, the
comparison is meaningless, since the run time is dominated by other effects.

Selecting a good sampling frequency is also a difficult problem and it might be
necessary to choose an adaptive strategy. Especially if the distribution of
types is skewed such that some types occur very infrequently, the profiler might
suggest wrong speculations and we must have a mechanism in place to detect and
prevent them from repeating. For instance, one example we found in the benchmarks
suite is the following snippet from the {\ttfamily nbody\_2} benchmark:
\begin{lstlisting}[%
  style=R]
drr <- array(dim=c(n_bodies, n_bodies, 3))
for (i in 1:n_bodies)
  for (j in 1:n_bodies)
    drr[i, j,] <- body_r[,i] - body_r[,j]
    ...
\end{lstlisting}

\noindent Here, the variable \rc{drr} is defined as an uninitialized array and then
subsequently updated with concrete values. The \rc{array} function from the
standard library initializes arrays with logical \rc{NA} values and therefore
the initial type of the array is of type {\ttfamily logical}. However the \rc{body_r} variable
is a double array, causing the first assignment to convert \rc{drr} to
{\ttfamily double} as well. When sampling the {\ttfamily nbody\_2} benchmark in the default
configuration, less than one in ten samples of that function contain the initial
{\ttfamily logical} type. On the other hand, this particular problem can also be solved by
improving type inference. In this case adding a single rule to the type
inference pass of the compiler is sufficient to statically infer the type of
\rc{array} when the initial value is left out. In general, we observed type
inference to be very robust at discarding or complementing implausible type
feedback. In particular, in combination with sampling-based profiling, where the
feedback can be sparse with many missing entries, we believe that an inference
pass to infer the most likely type could further
improve our results.

For a few benchmarks we have observed the profiler to incur a substantial
overhead. This overhead could be further reduced by running the profiler in
bursts.
We are also interested in exploring the use of profiles which are simply too
expensive to gather by instrumentation. For example, we would like to evaluate
the use of context sensitive profiling proposed by \citet{wha00} in our context.
Context sensitive profiling would allow us to detect candidates for
splitting, where different callers should be presented duplicates of their
common callee, hence avoiding false sharing of type feedback.

\section{Conclusion}\label{sec:conclusion}

We present a sampling-based profiler for a virtual machine, capable of monitoring native code without
instrumentation, used to detect inefficient code with missed optimization
opportunities. The absence of instrumentation has the advantage that
when the profiler is not engaged, no additional overhead is incurred.
We evaluate the approach on an implementation for the \R research virtual machine for the R language.
Our profiler uses the Performance Monitoring Unit (PMU) to trigger samples,
allowing to reliably interrupt the program only when running in user-level
code and not during syscalls.

Preliminary measurements indicate that it is possible to accurately detect stale or too
generic specialization in native code and improve peak performance by
recompilation with subsequent specialization to the sampled information.
The profiling was found to incur small overheads of typically below 3\% and up
to 13\% for some benchmarks and configurations. The main trade-off observed
concerns the recompilation threshold, that needs to be tuned to trigger as early
as possible while avoiding the use of incomplete data leading to wrong
speculation. Our evaluation considers many combinations of sampling intervals
and recompilation thresholds, which allowed us to determine a sweet-spot for our
implementation. Three short running benchmarks, which are representative for real-world situations with stale
type feedback, show encouraging improvements of 1.1-1.5$\times$.

It still remains to be seen if the proposed technique can be tuned and made
robust enough to improve the performance of real world R programs without
causing unexpected performance behavior. As future work, we proposed to explore
more robust heuristics, sampling of more properties besides types and the
recording of context sensitive samples.

\begin{acks}
We greatly thank all the people contributing to \R over the years, in particular
Guido Chari,
Jakob Hain,
Paley Li,
Petr Maj,
Ming-Ho Yee.
Thank you Oscar Nierstrasz, this work would not have come together if it weren't for
your passionate mentoring and teaching.
This work received funding from the Office of Naval Research (ONR)
award 503353, the European Research Council under the European Union’s
Horizon 2020 research and innovation programme (grant agreement
695412), the NSF (award 1759736), and the Czech
Ministry of Education, Youth and Sports (grant agreement
CZ.02.1.010.00.015\_0030000421).
\end{acks}

\bibliography{bib/pir.bib,bib/jv.bib}

\appendix\section{Appendix}\label{sec:appendix}
One large obstacle in the works presented in this paper was the correct configuration of the PMU.
We found the available documentation to be somewhat lacking in detail about achieving specific
configurations. In \autoref{lst:pmu} we present the configuration we used.

The first part configures a function {\ttfamily handler} (not shown) as the signal handler
for the {\ttfamily SIGUSR1} signal. Next, we prepare the PMU configuration.
We configure the PMU to count retired hardware instructions and to sample the instruction pointer at
a set period. In this configuration the PMU reads the current instruction pointer after the set number
of instructions. We also exclude the kernel and hypervisor. This prevents the PMU from taking a sample
while executing in these two modes. The {\ttfamily precise\_ip} flag set to $3$ forces the PMU to take
the sample immediately when the instruction counter reached its set limit. These two configurations
work together to prevent signal produced by the sampling from interfering with system calls.

Applying these configurations using {\ttfamily perf\_event\_open} returns a file
descriptor.
The samples taken by the PMU are written to a buffer associated with that descriptor. Next, we take
that file descriptor and configure it using {\ttfamily fcntl} to send a {\ttfamily SIGUSR1} signal
whenever new data becomes available (\ie a new sample has been taken).

Finally we reset the PMU and refresh the sample limit. Setting it to a value of $-1$ allows the PMU
to take the maximum number of samples (usually $2^{40}$). By using this, we get a basically
unlimited number of samples and don't have to refresh again.

\begin{lstlisting}[language=C,
  basicstyle=\small\ttfamily,
  belowcaptionskip=0pt,
  caption={PMU settings for repeatedly triggering signals},
  captionpos=t,
  label={lst:pmu}]
int signal = SIGUSR1;
// Register a signal handler
struct sigaction sa;
memset(&sa, 0, sizeof(sa));
// register handler function
sa.sa_handler = handler;
sa.sa_flags = 0;
if (sigaction(signal, &sa, NULL) != 0)
  exit(FAIL);

// Configure PMU
struct perf_event_attr pe;
memset(&pe, 0, sizeof(pe));
pe.type = PERF_TYPE_HARDWARE;
pe.size = sizeof(pe);
// Count retired hardware instructions
pe.config = PERF_COUNT_HW_INSTRUCTIONS;
pe.disabled = 1;
pe.sample_type = PERF_SAMPLE_IP;
pe.sample_period = 1000000;
// excluding kernel-space events
pe.exclude_kernel = 1;
// excluding hypervisor events
pe.exclude_hv = 1;
// precise event triggering
pe.precise_ip = 3;

// configure PMU for all threads of the current process and all CPUs
int fd = perf_event_open(&pe, 0, -1, -1, 0);
if (fd != 0)
    exit(FAIL);

// Connect perf event fd to signal
fcntl(fd, F_SETFL, O_NONBLOCK | FASYNC);
fcntl(fd, F_SETSIG, signal);
fcntl(fd, F_SETOWN, getpid());

// Reset event counter to 0
ioctl(fd, PERF_EVENT_IOC_RESET, 0);
// Allow first signal
ioctl(fd, PERF_EVENT_IOC_REFRESH, -1);
\end{lstlisting}

\end{document}